# Construction of Nano-Assembled Microcapsules Embedded with Gold Nanoparticles for use in Novel Drug Delivery Systems

By

Abraham Samuel Finny

This thesis is submitted in partial fulfilment of the requirements for the Degree of Master of Science in Biomedical Engineering

Dec 2014

# Copyright Statement

This thesis is the result of the author's original research. It has been composed by the author and has not been previously submitted for examination which has led to the award of a degree.

The copyright of this thesis belongs to the author under the terms of the United Kingdom Copyright Acts as qualified by University of Strathclyde Regulation 3.50.

Due acknowledgement must always be made of the use of any material contained in, or derived from, this thesis.

Signed : 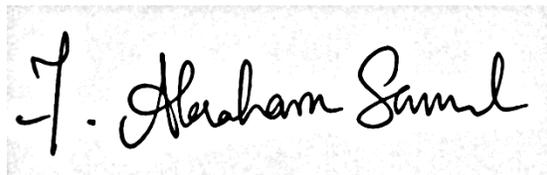

Date : 01/12/2014

*In loving memory of my Father, my first teacher and friend.*

*This one's for you Daddy!*

# Abstract

Coronary stents have changed the way in which coronary artery diseases are managed. Although bare-metal stents can be traced back to 1994, long term efficacy of these stents has been shattered by ISR (In-Stent restenosis) and late stent thrombosis [1]. Research on finding a solution to these issues has led to the development of DESs (Drug eluting stents). However long term effectiveness of DESs and various drug delivery systems have raised concerns. Also current DESs does not have the capability to adjust the drug dose or release kinetics that corresponds to the diseased status of the affected vessel. Through the use of a drug delivery system which employs controlled drug release, it may be possible to control the release rate of the pharmacological drugs or substances of interest, and therefore one might be able to circumvent the range of events that leads to ISR, thus preventing the need for further invasive interventions [2]. To overcome the limitations of drug delivery systems, the administration of drugs using nanoparticle based microcapsules as carriers is being researched for their ability to facilitate a sustained, prolonged and controlled release of drugs. But still much research is needed to evaluate and fix issues relating toxicity, chemical and mechanical properties of the nanoparticles [3]. This project combines a range of nanobiotechnological techniques to synthesise nano-components which are finally assembled together to get the microcapsules which could be used in drug delivery systems. The microcapsules were engineered keeping in mind their potential use in an ultrasound based drug delivery system.

## Acknowledgments

First and foremost, I would like to thank God for His grace and for providing me with the wisdom and strength to complete this research thesis.

I am indebted to a lot of good people without whose support, this project would have been nothing but a dream. I would like to thank my project supervisor, Dr. Wei Yao for his expertise and patient guidance throughout the course of the project and motivating me throughout the project till its successful completion. I would also like to express my gratitude and regards to Prof. Helen Grant for her moral support during my hard times.

I extend my appreciation to the members of the photophysics research group of the physics department for taking me in as their own and teaching me their secrets.
Especially Dr. Yu Chen, whose skills in the area of nanoparticles was indispensable. Also I would like to hold responsible the PhD students of the Photophysics research group especially Philip Yip for helping me with instrumentation and further motivating me to continue researching, and Peng Gu for his help with the chemical procedures and many other PhD candidates who were an inspiration.

I also thank the staff and my friends of the 2013/2014 batch of the biomedical engineering department. I also express my thanks to my friends, family and relatives in India without whom none of this would have been possible. Ultimately I thank everyone who had rendered help either directly or indirectly at various stages of the project.

*"I believe in intuition and inspiration. ... At times I feel certain I am right while not knowing the reason. When the eclipse of 1919 confirmed my intuition, I was not in the least surprised. In fact, I would have been astonished had it turned out otherwise. Imagination is more important than knowledge. For knowledge is limited, whereas imagination embraces the entire world, stimulating progress, giving birth to evolution. It is, strictly speaking, a real factor in scientific research."*

– Albert Einstein [4]

# Table of Contents

# Table of Figures

# Abbreviations

| | | |
|---|---|---|
| ACD | - | Atherosclerotic cardiovascular disease |
| ACE | - | Angiotensin converting enzyme |
| AHD | - | Atherosclerotic heart disease |
| AU | - | Absorbance units |
| BMS | - | Bare Metal Stents |
| CAD | - | Coronary artery disease |
| CCB | - | Calcium channel blockers |
| CHD | - | Coronary heart disease |
| CPB | - | Cardiopulmonary bypass |
| CPS | - | Counts per second |
| CVD | - | Cardiovascular disease |
| DES | - | Drug eluting stent |
| IHD | - | Ischemic heart disease |
| ISR | - | In-stent restenosis |
| LDL | - | Low density lipoprotein |
| MIDCAB | - | Minimally Invasive Direct Coronary Artery Bypass |
| NCD | - | Non-communicable disease |
| OPCAB | - | Off pump coronary artery bypass |
| PAH | - | Polyallylamine hydrochloride |
| PCI | - | Percutaneous coronary intervention |
| PDLA | - | Poly (D, L-Lactide) |
| PLA | - | Polylactide |
| PE | - | Polyelectrolyte |
| PSS | - | Polystyrene sulfonate |
| Rh6G | - | Rhodamine 6G |
| RPM | - | Revolutions per minute |
| SEM | - | Scanning electron microscope |
| SMC | - | Smooth muscle cell |
| SS | - | Stainless steel |
| WHO | - | World health organisation |

# 1. Introduction

According to data published by the world health organisation (WHO), out of the total 57 million deaths that occurred in the year 2008, 63% were due to non-communicable diseases (NCDs). A towering 48% (17 million) of these deaths due to NCDs were found to be due to cardiovascular diseases (CVDs), which is almost 30% of all global deaths [5]. And of these 17 million deaths, approximately 42.2% (7.3 million) were attributed to coronary heart disease (CHD) also known as coronary artery disease (CAD) [6]. Statistical data also suggest that the number of deaths due to CVDs is projected to reach an overwhelming 23.3 million by 2030 [7]. Therefore, the need to research and develop novel methods to treat this global problem is paramount.

CHD, which is the most common CVD, is caused due the narrowing of the lumen of the arteries in the heart which occurs due to the build-up of plaque along the inner walls of the arteries, which causes reduced blood flow thus depriving the heart of oxygen and nutrients to the cardiac tissues. One of the main treatment of this narrowing of the arteries within the heart is by the placement of stents in the narrowed region to bring back the normal blood flow. But the main issue with this 'stenting' is the high rate of occurrence of in-stent-restenosis. In order to deal with this, drug eluting stents (DESs) are being developed and used. A DES is usually made up of three parts: the body of the stent which provides the structural support, the drug to be eluted and finally the drug carrier [8], [9], [10].

The drug carrier in the stent is a significant component in DESs. The quality and the effectiveness of a DES directly depends on the drug carrier. They should meet a wide range of requirements before being able to be used in stents. The drug carrier (usually a polymer) should be biocompatible in order to prevent any inflammatory/rejection reactions. Also one should ideally be able to control the drug release therefore the concentration of the drug within the arterial wall can remain within the therapeutic window. This would enable the drug concentration to be high enough to hamper the proliferation of smooth muscle cells (SMCs) in a timely manner without causing any short/long-term effects which may adversely affect the arterial walls [11].

Recent developments in biomedical engineering have made the development of newer and better drug delivery systems for stents possible, especially in combination with advanced nanotechnological techniques. Novel drug delivery mechanisms which utilise drug encapsulation, followed by drug release through an external stimulus (e.g. ultrasound) which can be triggered once the encapsulated drug is at the site of interest has gained a lot of interest over the last years [12]. Clinical usage of these kinds of new novel drug delivery systems for treatment of in-stent restenosis are however in their early stages and this area of research needs further attention as the possibilities are vast. This project work therefore explores this area in an attempt to create a drug carrier which can be used in an advanced drug delivery system, preferably for a drug eluting stent and that which could be possibly used in a drug delivery system which employs ultrasound for drug release.

## 2. Aims and Objectives:

The main aim of this thesis was to explore a new concept which would enable the development of novel drug delivery systems and contribute to research in the area. Utilising previous research in this area, this project experiments in creating a generic drug delivery capsule using recent nanotechnological processes and photophysical techniques. The objective was to obtain nano-assembled microcapsules which could be used in various drug delivery systems. The long term goal of the project was to use the obtained microcapsules embedded with gold nanoparticles in novel drug delivery systems. The ingenuousness of the method explored here is that if the process to make the microcapsule was to prove successful, the method could be used to get similar drug carriers which could be put into a wide range of uses. However, one of the chief motives of this experiment was to produce a system which could deliver drugs for the treatment of CHD, specifically for ISR. Another motive being the ability to control the elution of the drugs from the system through non-invasive means. Since ultrasound is non-invasive, future research which focusses on drug elution from these carriers using ultrasound would prove very beneficial to drug delivery research. By exploring previous and current research and incorporating innovative techniques, the design of an advanced drug delivery system would be possible. But such a drug delivery system would need advanced drug carriers which should be dynamic and efficient. This project strived to create such a drug carrier. The microcapsules obtained could also be coated onto DESs and thus provide cardiologists the world over with the ability to control drug release and one-day systemic drug administration could be a thing of the past.

## 3. Review of Literature

Biomedical engineering has seen tremendous growth over the past decade. This growth has led to the design, development and research of novel medical devices and systems which has contributed much to the field of healthcare. However, next generation medical devices and systems are always in demand, and in order to further advance the quality and level of treatment that a patient can get, innovation in this area is a necessity. In order to design any medical system, one must be familiar with the disease etiology and also with the functioning of the components of the said system. The section below hopes to address the current clinical options, their limitations and the need for the development of a novel drug delivery system for use in CHD treatment.

### 3.1 Coronary Heart Disease

CHD is also known as coronary artery disease (CAD), atherosclerotic cardiovascular disease (ACD), atherosclerotic heart disease (AHD) or ischemic heart disease (IHD) [13], [14], [15], [16], [17]. CHD affects coronary circulation thereby directly affecting blood supply to the heart. Coronary arteries are the vessels which deliver blood that is rich in oxygen to the myocardium. But these arteries are susceptible to atherosclerosis. The process of atherosclerosis is not yet fully understood, however research has found that it is initiated due to inflammatory responses by the endothelial cells of the arterial walls in response to low density lipoprotein (LDL) particles [18]. The disease starts with the formation of fatty streaks which are made up of LDLs and white blood cell deposits (both alive and dead cells) as seen in the image (Figure 1) below. This disease tends to progress very slowly, generally over decades and mostly remains asymptomatic until the accumulated atheroma breaks away, which may lead to a blood clot at the site of the atheroma rupture, or the ruptured atheroma may form a clot at a narrower region in the artery. The above events may lead to a complete blockage of blood flow which leads to ischaemia of the myocardium thereby triggering a myocardial infarction which can prove fatal [19].

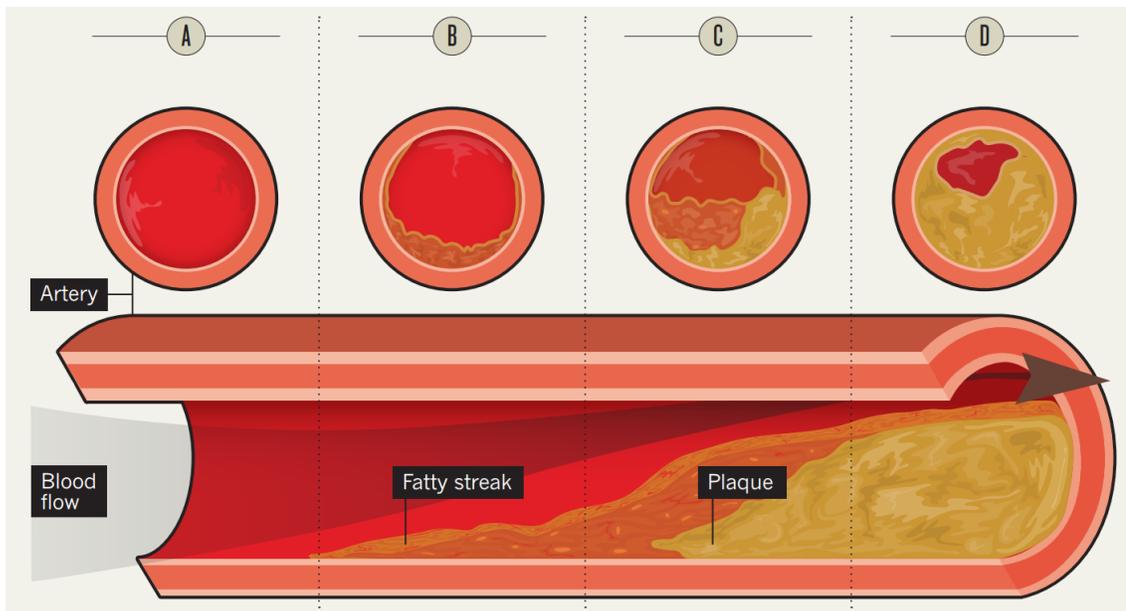

**Figure 1**: Atherosclerosis; A) Healthy arteries. B) Initiation of atherosclerosis (Fatty streak). C) Plaque formation. D) Pathway to a clot [20].

Though there are a wide range of reasons which may lead to stenosis in the arteries, so far no one has been able to pinpoint any one causative process that is definitive. However recent research has been able to put together a range of processes such as thrombosis, inflammation, proliferation, remodelling which may precede the condition [21].

### 3.1.1 Treatments

CHD treatments are usually dependent on the symptoms exhibited by the patient and the severity and/or the stage of the disease. Usually the treatment options for a CHD may be categorised into three types. 1) Through drugs 2) Through coronary artery bypass grafting (CABG) 3) Through coronary interventions [22]. Non-medical treatments such as lifestyle changes has also been proved to be effective in reducing the occurrence of CHD. For example, diet changes in some cases have even proved to reverse CHD [23]. Therefore, lifestyle changes are advised strongly for persons who are CHD prone and for patients who already have heart diseases. However, if one was to be diagnosed with CHD, lifestyle changes alone would not be enough and therefore suitable medical treatments are necessary to treat the problem. This medical treatment could however be assisted by positive lifestyle changes to aid in recovery and improved efficacy of the treatment.

### 3.1.1.1 Treatment through Drugs

Early stage atherosclerosis is usually treated through administering pharmaceutical drugs. The purpose of these drugs are to prevent further complications and to ease the symptoms for the patients. For example, statins can be used to reduce the amount to LDLs circulating in the blood; to prevent the reoccurrence of a myocardial infarction due to atherosclerosis, angiotensin converting enzyme (ACE) inhibitors and beta blockers can be used. Most of the time, a combination of drugs is used to treat the disease. Systemic drug administration is still one of the main ways used to treat atherosclerosis.

Pharmaceutical agents such as lipid-lowering drugs, anti-inflammatory drugs, antibiotics, antioxidants, antiproliferative drugs and other similar drugs hold good potential to alter the natural process of atherosclerosis. Drugs which interfere with the platelet-vessel interaction, inflammatory response, matrix production and proliferation of cells have also shown good results in vitro [24]. Angina can be managed using nitrates, calcium channel blockers can be used to reduce the heart rate and also to dilate the blood vessels [25] while aspirin can be used to reduce clot formation tendencies. However, usage of drugs at a later stage of the disease can be ineffective at times and therefore invasive methods of treatment are needed.

**3.1.1.2 Treatment through CABG**

CABG is more effective than medical management through drug administration at relieving symptoms of CHD [26]. Usually CABG is carried out on patients who are at a higher risk to have a heart failure due to CHD. This is a type of arterial revascularization in which the coronary circulation to the myocardium is re-established by means of creating a connection which bypasses the blocked blood vessel which in this case would be an artery. The connecting vessel usually would be ideally taken from somewhere else from the patient's body to maintain biocompatibility and to prevent graft rejection. This treatment was successfully introduced in 1960 by Robert H. Goetz who was a surgeon and his team [27]. This procedure has rapidly become a standard treatment for people with advanced stage CHDs over the years.

Modern technology has enabled the development and further improvements of minimally invasive CABG. Off pump coronary artery bypass (OPCAB) for example, is a technique of performing CABG without using CPB (Cardiopulmonary bypass - commonly known as the heart-lung machine) [28]. Newer innovations to OPCAB have resulted in a further minimal invasive treatment method known as minimally invasive direct coronary artery bypass surgery (MIDCAB) also known as a "keyhole" heart surgery which is a technique of performing CABG through an incision of 5-10cm size [29]. The incision is usually made between the ribs and entry to the chest cavity is gained by a minute thoracotomy. The treatment and recovery times of these newer methods are far less than the conventional CABG.

### 3.1.1.3 Treatment through Coronary Interventions

Percutaneous coronary intervention (PCI) also known as coronary angioplasty or balloon angioplasty is a non-surgical form of CHD treatment. PCI involves the insertion of a deflated balloon or a suitable device on a catheter through the femoral or radial artery to reach the site of blockage in the heart. To guide the catheter to the site of blockage, X-ray imaging techniques are used. Once at the site, the balloon is inflated to open up the block thereby restoring blood flow. When the balloon catheter is removed from the site of blockage, the lumen diameter would to be expanded, however due to elastic recoil, the risk of a blockage after the procedure became much evident. Also the occurrence of a restenosis after a certain amount of time, which resulted in the loss of the luminal area is still a cause for concern [30]. To address these issues, a coronary stent is introduced into the blood vessel to keep it open after the procedure, the stents could be coated with drugs to inhibit restenosis. PCI in combination with stenting is preferred for non-severe stages of CHD [31]. PCI has also shown consistently its ability to reduce symptoms due to CHD and also to lessen cardiac ischemia [32].

**3.2 Coronary Stents**

A coronary stent is a flexible mesh-tube like structure which is inserted by means of catheterisation into a coronary artery to keep it open. It is an arterial wall support structure which is designed to stay inside the artery [33]. Stents are widely used these days in PCI to either restore or maintain blood flow to the myocardium [34]. Stents can be broadly classified as inert (non-eluting) and drug eluting stents (DESs) [35]. At present most of PCI procedures are assisted by stenting. Stenting procedure follows similar steps as that of PCI. The stent is mounted in a catheter in a compact form and on reaching the site of the block, it is deployed. Usually deployment is done through means of balloon expansion. For over 20 years this stenting technique has been in use for treating CHD [36]. An image explaining stent deployment can be found in figure 2.

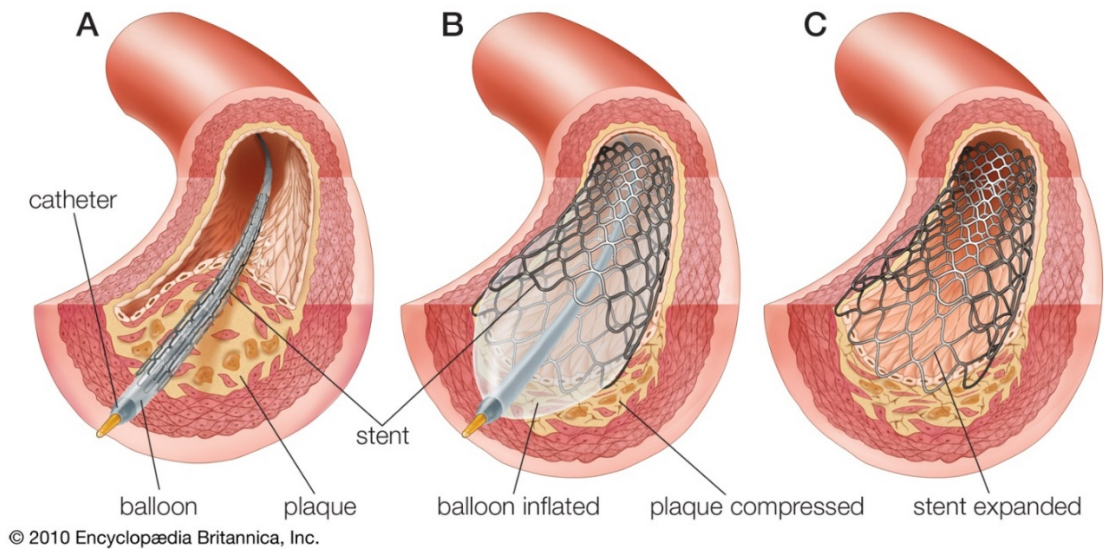

**Figure 2**: Stent deployment; A) Catheterization B) Balloon inflation
C) Balloon deflation and stent deployment [37].

Studies show that stents have reduced the occurrence of restenosis after PCI treatments in comparison to treatments involving balloon angioplasty without stenting. This is achieved because the stent opposes elastic recoil and late remodelling of the artery [38], [39]. Still, the excess neointimal hyperplasia inside the stent induces a restenosis in 10%-30% of the patients who are treated [40]. This in-stent restenosis (ISR) which is a local and chronic response is a major issue which affects the treatment of CHD.

### 3.2.1 Coronary Stent Design

There are a wide range of factors which should be taken into account while designing a stent for use in the coronary arteries. The foremost thing to consider is the biocompatibility of the material which is to be used for the stent construction. The material properties of the stent also play a major role in the performance of the stents. Mechanical properties such as surface smoothness, corrosiveness, strength, elasticity, ductility, etc. are also vital. The degradation of the material due to corrosion inside the artery can lead to the leeching of metal ions which can be toxic and therefore can affect the arterial wall or the tissues nearby [41]. Some of the commonly used materials for the construction of metallic coronary stents are 316L SS, Pt-Ir alloys, Ta, Ti, Ni-Ti, and Co-Cr alloys. For biodegradable metallic coronary stents, pure Fe and Mg alloys are used commonly [42]. 316L SS is the most commonly used material and it provides good mechanical properties. It is also corrosion resistant. Other unconventional materials used in the construction of stents include nitinol which is a shape memory alloy, platinum, etc.

When the materials which are used have a high radial strength, the struts can be made thinner without compromising structural support. Poor radial strength materials lead to the struts of the stents being thicker. It is very important that the stent has enough radial strength to keep the vessel open without collapsing on itself. The flexibility of the stent enables it to be guided properly by the catheter without affecting the blood vessels. Usually the thinner a stent is, the more flexible it would be.

When using bare metal stents, biocompatibility and hemocompatibility still remains a large issue. Thrombosis and neointimal hyperplasia were found to occur commonly over the stented area [43]. Therefore coating the surface area of the stent with other materials to change the surface properties without affecting the mechanical properties of the stents has become a common method to address the above issues [44], [45]. Metallic coatings are the most common coatings used in stents. Recent advances in coating stents have made it possible for the use of polymers. Polymers that are used for coating coronary stents can be categorised as non-biodegradable, biodegradable, copolymers and biopolymers. Although there are a wide range of polymers which can be used to coat stents a few such as Polyethylene terephthalate (PET) and a few forms of polylactide (PLA) have been proved successful [42]. Since a lot of research exists on the feasibility of using PLA for stent coatings, this project uses PLA as the core for the microcapsule, hoping that the previous research could justify the selection of PLA as the material of choice.

**3.2.2 In-Stent Restenosis**

Almost half of the population of the patients who have had bare metal stents implanted had faced the occurrence of stenosis within the stents [46]. ISR can be mainly attributed to viral neointimal proliferation. However recent studies show that neoatherosclerosis may also play a vital role in the pathophysiology of ISR [47]. ISR can usually be said to have occurred if the diameter of the recurrent stenosis is greater than 50% of the diameter of the stent or over the edges of the stent [48], [49]. ISR can also be traced to stent fractures [50] and to injuries that occurs while stenting.

The image (Figure 3) below graphically explains the various processes which occurs after a stent-induced arterial injury. Stent-induced arterial injury is one of the precurosors of ISR.

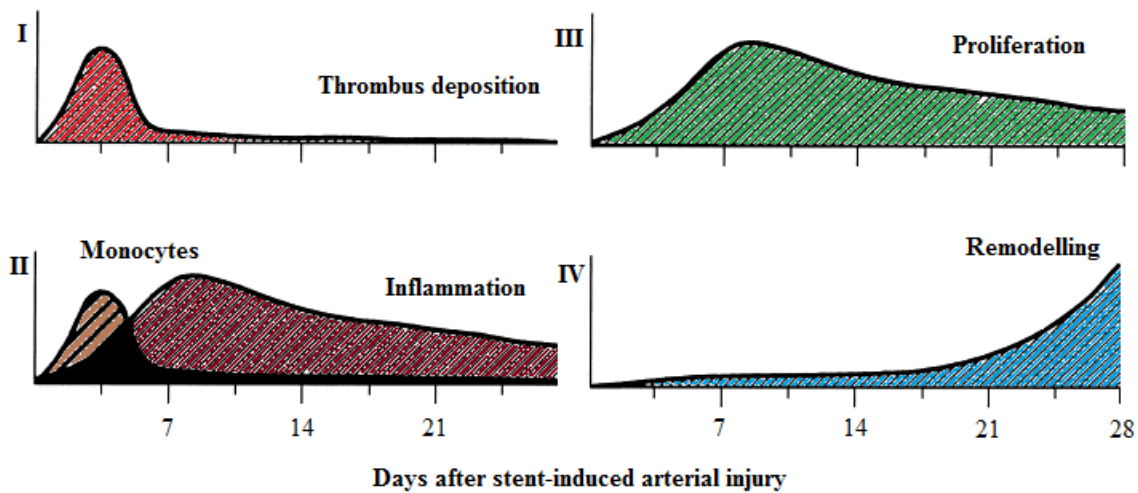

**Figure 3**: Initiation of ISR after stenting; I) Accumulation of platelet rich thrombus II) Arrival of monocytes at the site of inflammation III) Proliferation of SMCs and monocytes IV) Arterial shrinkage [51].

Administration of drugs like heparin, ACE inhibitors, nitric oxide donors, abciximab, hirudin, tranilast, angiopeptin in a systematic manner had shown good results in laboratory over animal studies for ISR treatment but have failed in clinical trials [52], [53]. The above failure was probably due to the inefficient concentrations of the said drugs locally and also can be blamed partially on the poor knowledge of the pathophysiology of restenosis at the time. This problem is in the process of being addressed by coating stents with suitable pharmacological agents on the surface thereby inhibiting restenosis [54]. Early studies dating back to the 1990s which used polymer coated stents for the treatment and prevention of restenosis has shown positive results [55]– [57]. As previously mentioned, a range of oral therapeutics like ACE inhibitors, statins, calcium channel blockers (CCBs), vitamins, antiproliferative, antiplatelet and anticoagulant drugs have also showed promising results. Out of these drugs, the most effective drugs at addressing ISR were found to aspirin and heparin which are antiplatelet and antithrombotic drugs respectively. Systematic administration of the said drugs decide whether the treatment is a success or not [58].

### 3.2.3 Drug Eluting Stents

Even though stenting has reduced the occurrence of restenosis in comparison to treatment that includes balloon angioplasty alone, the very high rate of occurrence of ISR has led to increased interest in DESs. It is expected that DESs would be the route to solve the problem of ISR [59]. CHD treatment has been revolutionised over the past decade through the use of DESs [36]. Also the arrival of DESs has also not only reduced the occurrence of ISR, but has also made it possible to treat more complex lesions without the need of a much more invasive bypass surgery [21]. The need for usage of DESs can be attributed to the fact that oral drug administration does not have an effective impact on the rate of restenosis [60]. The main reason for the above maybe due to the inadequate concentrations of the drug at the stent site, which can be addressed through the use of DESs. DESs so far has had a huge effect on the rates of ISR; 30 to 40% of ISR occurrence in patients with BMS compared to 0-9% with patients with DES [61]– [64]. A strut of a typical drug eluting stent can be seen from the image (Figure 4) below.

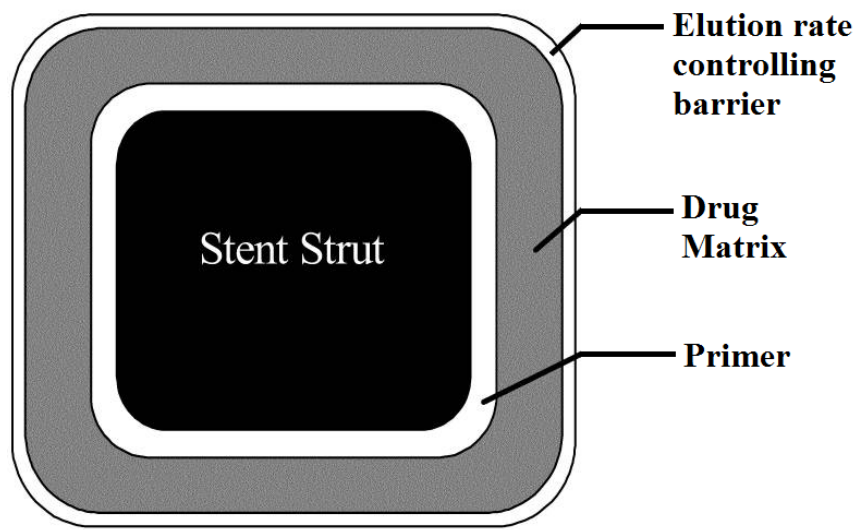

**Figure 4**: Cross sectional view of the strut of a generic DES [65].

**3.2.3.1 DES' Drug Delivery Mechanisms**

The drugs in DESs can be released either by physical or chemical means. Physical means of altering drug elution rates are preferred over chemical release mechanisms because for the latter, the carrier would need to carry modified drugs which would result in extra chemical entities [66]. The mechanisms of drug release can be further classified into five systems viz. Diffusion system, degradation system, ion exchange system, osmosis system and prodrug system. Diffusion and degradation systems can be differentiated into reservoir and matrix sub systems [66]. This thesis aims to combine a few of the above systems into one in an attempt create a novel mechanism for drug delivery.

### 3.2.3.2 Limitations of DES

Technologically DES are far more advanced than bare metal stents, however DESs do have limitations that need to be addressed. Many DES studies which show promising results do not take into account many complex real world lesions. Also the results of studies which included more complicated lesions suggest that even though occurrence of ISR might have been reduced, they have not been completely eradicated [67]– [69]. For most of the DESs, studies show that the rate of drug release over the first few days of stent implantation is very high and hence most of the elution happens within that period while in some cases drug retention within the coating is also a cause for concern [70]. Therefore, it is highly necessary to create a delivery system where the drug release can be controlled.

The other major limitation is the injury due to the stent itself. When any stent is placed in an artery, the chances of damage to the endothelial layer is almost certain. When the endothelial layer is affected, ISR could be initiated as a result of the various inflammatory responses. ISR can be observed in BMS more easily because the onset is early, however in DES the detection can be late as the onset of ISR delayed which is a problem [71]. Late stent thrombosis is another major issue that haunts DESs. Late stent thrombosis is the formation of a clot at the stent site which can cause the artery to occlude in the longer run. Late stent thrombosis as the name suggests does not occur immediately but rather years after the stent has been implanted. This makes the situation complex in case of old patients who would have aged further when this complication arises, thus making further treatments tricky.  In order to overcome these limitations, designing a drug delivery system which can elute and retain drugs as and when required while remaining stable within one's body over the years is imperative.

## 3.3 Modern Drug Delivery Systems

Recent technological advances have led to the engineering of better drug delivery systems. The use of nanotechnological techniques in biomedical engineering has led to a wide range of innovations which include better drug delivery systems [72]. Nanotechnology in this aspect includes the production and assembly of substances at the nanoscale including characterization and analysis using devices and also encompasses nanobiotechnology [73]. Nanomedicine in the present day has given rise to advanced drug delivery techniques [74]. Drug encapsulation in the nanoscale which aims to improve the efficacy, biodistribution, bioavailability of the drug while minimising the problems with drug concentration that are common in systemic drug administration, is one of the most promising areas in nanotechnology in modern times which offers much potential [75]. The figure 5 below represents diagrammatically some of the commonly used drug carriers.

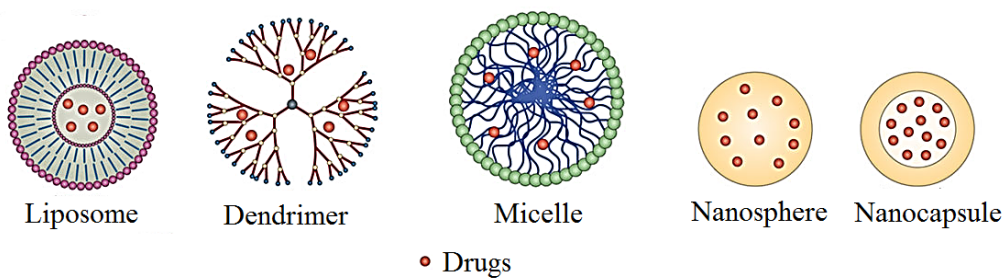

**Figure 5**: Various novel nanocarriers [76].

Encapsulating drugs by means of nanocapsules as visualised in the figure above or using microcapsules and using them for drug delivery has become an exciting area to research because of the various possibilities it has to offer. Liposomes, nanospheres, nanocapsules, polymer conjugates, nanotubes, cells and various nanoparticles have shown the abilities to be used as nanocarriers. Microencapsulation offers protection for the encapsulated drug, makes it possible to control the drug release, makes it easier to administer the drug and finally provides one with the ability to pre-programme release profiles to match the needs [77]. Also the capsules have the potential to be engineered in such a way as to respond to stimuli such as pH, ultrasound, temperature, specific chemicals and radiation [78].

**3.3.1 PE Microcapsules**

One of the ideal encapsulation methods is through the use of poly electrolyte (PE) microcapsules. Although they are termed as capsules, they are not capsules per se, but rather dispersed solid matrices lacking a definitive external wall phase. PEs are polymeric molecules which are obtained from ionisable monomers. PEs are also charged when suspended in an aqueous phase. PE microcapsules are made by coating differently charged PEs in alternative layers. PEs could also be used to coat colloidal particles which makes them suitable for use in a drug delivery system [12]. Maximum sizes of PE microcapsules can reach around 2000 micrometres depending on the layer thickness and the number or layer coatings. One advantage of using PEs is being able to modify the permeability of the capsule layer [79], thereby drug release and elution can be controlled. In PE microcapsules, another advantage is the possibility of being able to coat the drugs between the layers instead of just encapsulation, this enables the usage of multiple drug layers and provides sustained drug release scenarios [80]. One of the most efficient processes of synthesis of microcapsule is through layer by layer assembly. This is achieved through the use of poly cations and poly anions. The most common poly cation - poly anion pair, PAH (Polyallylamine hydrochloride) and PSS (Polystyrene sulfonate) has proved to be among the best combination for creating PE microcapsules [78]. This layer by layer assembly is attained by coating one charged PE layer above the other PE layer that holds the opposite charge. This layering can be done as required [72].

### 3.3.2 Gold Nanoparticles

Embedding gold nanoparticles on the PE layers of the microcapsule would provide a lot of desired properties to be added to the microcapsule which aid in optimal drug delivery. The sensitivity of the microcapsules would also be considerably increased and therefore the use of any stimuli to trigger drug release would be made comparatively easier [78]. Also unwanted retention of drugs within the capsules could also be lowered significantly when microcapsules embedded with gold nanoparticle use a stimuli based release system like ultrasound [12]. In addition to these, gold is inert and non-toxic which further makes it a preferred material of choice [81]. Moreover, the gold nanoparticles play a role in improving the mechanical properties of the microcapsules by providing it with stability [82].

## 4. Materials and methods

### 4.1 Introduction

The experimental methodology of this project aims to show the potential that nanobiotechnology has to offer for the creation of a new drug delivery system. In the history of drug delivery systems, microcapsules that are loaded with drugs have been used for a wide range of therapeutic purposes. The following experiment researches the possibility and feasibility of using PLA (Polylactide) nanoparticles, gold nanoparticles and certain polymers (PEs) to create a novel drug delivery system. The main idea was to create a generic drug carrier which would be able to encapsulate drugs of interest to treat CHD. The intention was to dope a nanoparticle core which was made of PLA with Rhodamine 6G dye. Rhodamine 6G was used as a substitute in place of the pharmaceutical drug that has anti-ISR properties. Also Rhodamine has a range of photophysical properties and hence photophysical measurements could also be taken. Thus the PLA core infused with Rhodamine 6G would be made to undergo a multiple layering process, using alternative charges through employing Polyallylamine hydrochloride (PAH) and Polystyrene sulfonate (PSS) thereby encapsulating the PLA core and forming microcapsules. Therefore, the PLA core would be first coated with a layer of PAH, followed by a layer of PSS. The third layer would be coated using gold nanoparticles and finally another layer of PAH can be used to form another layer. Figure 6 clearly represents the coating process visually.

Since the layering is based on charge difference, the final layer attains a negative charge in our case because of the gold nanospheres. This offers us the possibility to coat it on stents by positively charging the stent strut.

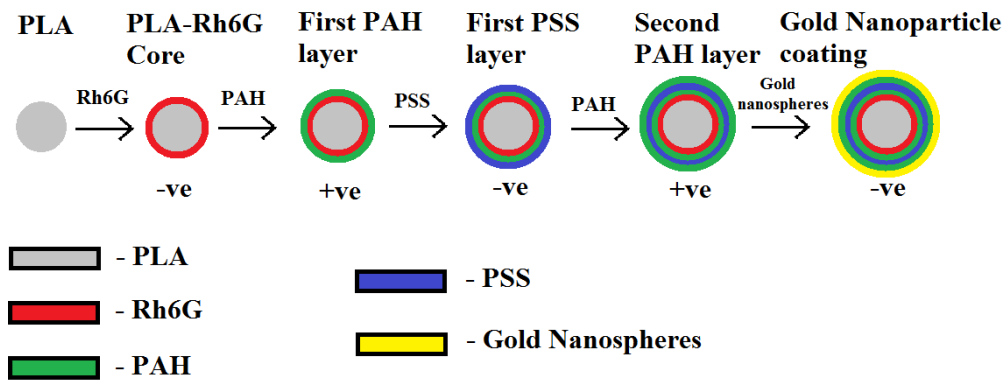

**Figure 6**: Diagrammatic representation of the layering procedure (Not to scale).

**4.2 Polylactide**

Polylactide is an aliphatic polyester which is also a biocompatible thermoplastic and has the potential to be used for controlled drug releases [83]. It is also biodegradable and lipophilic and has been used extensively in the biomedical field [84]. Figure 7 shows the structure of DL-Lactic acid.

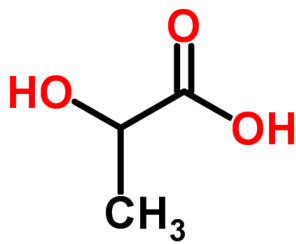

**Figure 7**: 2-D chemical structure of DL-Lactic acid.

The extensive use of PLA in the field of biomedical engineering can be attributed to its good mechanical properties and also to its ability to degrade both in vivo and in vitro to non-toxic products [85]. PLA is also nontoxic inside the human system which furthermore makes it suitable for use in several medical applications [86].

**4.3 Polyallylamine Hydrochloride and Polystyrene Sulfonate**

PAH is a cationic polyelectrolyte (PE) obtained through the polymerization of allylamine. PSS is an anionic polyelectrolyte derived from polystyrene. The above polymers can be combined together to form layer-by-layer adsorbed films of negative-positive charges polymers [87]. The most prominent biomedical application of the above polymers is in the field of encapsulation. The capsule of layered PEs has very good structural integrity and has the potential to be used for drug delivery purposes [88].

**4.4 Rhodamine 6G**

Rhodamine 6G (Rh6G) is a highly fluorescent dye. It is derived from xanthene and is frequently used an active medium and probe for photophysical analyses [89]. Rh6G is an ideal probe to be used because of its high light absorption and fluorescence properties [90], [91]. In this experiment Rh6G is used generically in the place of a drug as a substitute, because of its property to be monitored easily through photophysical means. The chemical structure of the dye is shown in the figure 8 below.

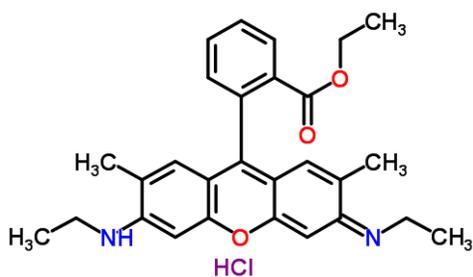

**Figure 8**: 2-D chemical structure of Rhodamine 6G.

Also Rh6G in combination with the biodegradable PLA can mimic certain characteristics of the real drug but also present one with the ability to monitor its elution. However, Rh6G cannot be used in real life trials or treatment because it is toxic and maybe even carcinogenic for humans.

**4.5 Gold Nanoparticles**

Gold nanoparticles have the ability to act as non-toxic nanocarriers which can be used for drug delivery applications. The gold nanoparticles provide stability to the capsule and the layer also allows surface properties such as hydrophobicity and charge variations. Also gold nanoparticles interact with thiols, thus enabling an effective and selective use for controlled intracellular release [82]. Recent research has made gold nanoparticles as an attractive tool for use in drug delivery systems [92], [93].

**4.6 Layering Process**

Assembly of the microcapsule will begin by taking the core doped with Rh6G which is negatively charged and coating it with positively charged PAH followed by negatively charged PSS followed by a second layer of PAH. The PAH and PSS combination is one of the most commonly used layering PE combo. Once the second PAH layer is complete, negatively charged gold nanoparticles would be used to form a layer above the positive charged PAH layer. The layering process can be done as many times as required to get the preferred number of layers. Many studies have been made to determine the ideal number of layers, however there is a lot more work needed to come to a solid conclusion on how the layers may affect the various characteristics of a microcapsule. In our case, the layering process is kept to a minimum; three PE layers followed by a gold nanosphere layer.

One of the important things that needs to be correctly achieved in the layering process is to give the particles between each layer the proper temperature and proper mixing time to get a stable layer. Though the PE layer coatings did not need any change in temperature, for embedding the gold nanoparticles in the final layer, the PE coated PLA particles were exposed to boiling temperatures. The concentration of the surfactant was another parameter that needed attention. It should be noted that at one point in the experiment a high concentration of citrate was added due to a calculation error and it was not possible to synthesise any gold nanoparticles at all. Concentration of citrate was crucial factor in synthesis. The sizes of the nano-components should also be optimal for the layering to succeed.

## 4.7 Syntheses

There were two main nano-components which were needed to be synthesised before they could be assembled; The PLA-Rh6G nanoparticles which were to be used as the core and the gold nanospheres which were to be embedded on the final layer of the microcapsule.

### 4.7.1 Synthesis of PLA-Rh6G Nanoparticles

#### 4.7.1.1 Chemicals

- Poly(D,L-lactide) [PDLA (A form of PLA), $((C_6H_8O_4)_n)$, ester terminated, $M_w$=10,000-18,000], Sigma Aldrich [94]
- Acetone (CHROMASOLV®) [$(CH_3COCH_3)$, $M_w$=58.08], Sigma Aldrich
- Rhodamine 6G [$C_{28}H_{31}N_2O_3Cl$, $M_w$=479.01], Sigma Aldrich
- Distilled water

#### 4.7.1.2 Instruments

- Analytical balance (Model: KERN ABT 120-5DM)
- Hotplate and stirrer (Model: Jenway 1000)
- Microfuge (Model: Sigma 1-14)
- Ultrasonic bath (Model: Fischer FB15051)
- Magnetic stirrer bars, centrifuge tubes, micropipettes, beakers, etc.

**4.7.1.3 Method**

The synthesis of the PLA core was carried out according to previous published research work [95] [96]. The infusion of Rh6G with the synthesized PLA core was also carried out according to published research, where Rh6G is added to the acetone phase of the synthesis process [97]. This is because Rh6G is lipophilic. The process was as mentioned below.

3.33 ml (ideal acetone-water ratio) of acetone was pipetted into a beaker. 5 mg of PDLA was then measured using the balance and added to the acetone solution. The solution was gently agitated to facilitate the dissolving of the PDLA. Now 0.1 mg of Rh6G was carefully measured and added to the above solution. The colour change of the colourless solution to an orange red colour due to the Rh6G dye was visualised. A magnetic stir bar was then added to the solution and the solution was placed on the stirrer for 1 to 2 hours to enable the Rh6G and PLA to attach. Once the stirring was over, the solution was pipetted out into 8.33 ml of distilled water under moderate stirring drop by drop. This is the aqueous phase.

Once the above process was complete, the solution was transferred into 1ml centrifuge tubes and centrifuged for 30 minutes at 4,500 rpm followed by 10 minutes at 10,000 rpm. The supernatant was taken and kept aside for photophysical analyses and the polymer pellets infused with Rh6G were transferred to 10 ml of distilled water and ultrasonicated to enable suspension in the water. This made the PLA-Rh6G particles to get suspended in the solution which was later used as the core of the microcapsules to be assembled.

### 4.7.2 Synthesis of Gold Nanospheres

#### 4.7.2.1 Chemicals

- Gold(III) chloride trihydrate [(HAuCl$_4$ · 3H$_2$O), M$_w$=393.83, 25mM]
- Sodium citrate dihydrate [(HOC(COONa)(CH$_2$COONa)$_2$ · 2H$_2$O), M$_w$=294.10
- Distilled water

#### 4.7.2.2 Instruments

- Analytical balance (Model: KERN ABT 120-5DM)
- Hotplate and stirrer (Model: Jenway 1000)
- Ultrasonic bath (Model: Fischer FB15051)
- Magnetic stirrer bars, micropipettes, beakers, etc.

#### 4.7.2.3 Method

To synthesise the gold nanospheres for use as nanoparticles in the experiment, the established method to form citrate stabilized gold nanoparticles [98] was followed. It was necessary that the nanoparticles which were synthesised would be nanospheres as the shape was vital to the success of the delivery system and also essential for the coating mechanism that was to be employed.

150 ml of distilled water was pipetted into a beaker. 0.034 g of sodium citrate dihydrate was measured and added to the beaker. The beaker was then covered to prevent evaporation and the contents were dissolved by ultrasonication by placing it in the ultrasonic bath for 2 to 3 minutes. To further dissolve the mixture, the beaker was placed on the hotplate and was brought to boil under stirring. Once the solution started to boil, 1ml of 25 mM gold chloride trihydrate was pipetted into the solution while it was being stirred. The solution was then taken off the stirrer. The colourless solution undergoes colour changes until it finally becomes ruby red after cooling down. This significant and characteristic colour change marked the success of the gold nanosphere synthesis as the colour matched the absorption spectrum of the nanoparticles of this size range. The size of the nanospheres obtained through this process was approximately 10 nm and their concentration in the solution was around $3 \times 10^{12}$ nanoparticles per ml. The charge of the nanospheres was negative because of the citrate ions.

### 4.7.3 Layer by Layer Assembly

#### 4.7.3.1 Chemicals

- Poly (allylamine hydrochloride) solution (1g/l)
- Poly (styrene sulfonic acid sodium salt) solution (1g/l)
- Distilled water
- Citrate stabilised gold nanoparticles (nanospheres)

#### 4.7.3.2 Instruments

- Analytical balance (Model: KERN ABT 120-5DM)
- Hotplate and stirrer (Model: Jenway 1000)
- Microfuge (Model: Sigma 1-14)
- Ultrasonic bath (Model: Fischer FB15051)
- Magnetic stirrer bars, centrifuge tubes, micropipettes, beakers, etc.

#### 4.7.3.3 Method

The PLA-Rh6G particles which were earlier synthesised and suspended in distilled water were now used for the assembly of the microcapsules. At each stage in the process, the supernatant was kept for photophysical analyses. The layering was carried out in a ratio of 1:1. The PLA-Rh6G was earlier suspended in 10ml of distilled water. This 10 ml was further concentrated to 5 ml through centrifugation and resuspension. After this, 5 ml of the PAH solution was added to the above and placed on the stirrer for 2 hours. The solution was then centrifuged for 30 minutes at 14000 rpm. After centrifugation the particles were separated and resuspended in 5 ml of distilled water through ultrasonication. After resuspension, 5 ml of PSS was added to the above and then placed on the stirrer for 2 hours. After the stirring process, the solution was centrifuged as before and the pellet resuspended in 5 ml of distilled water. A second PAH layer was then formed over the above PSS layer using the above method. And finally 2 ml of the gold nanosphere solution was added to the solution which had suspended particles of PLA coated with PAH-PSS-PAH. Again the solution was placed on the stirrer for 2 hours, followed by centrifugation and finally resuspension. This solution would have the microcapsules with gold nanospheres as the outermost layer. Another layer of PAH over the gold layer followed by PSS can also be produced. The ingenuity of this method is that the number of layers can be varied as required.

**4.8 Photophysical Analyses**

**4.8.1 Instruments**

- V-660 Research UV-Visible Spectrophotometer by Jasco
- FluoroLog®-3 Spectrofluorometer by Horiba Scientific

**4.8.2 Software**

- FluorEssence™ for Windows® by Horiba Scientific
- SpecWin Pro
- OriginPro™ by OriginLab®
- Microsoft® Excel

**4.8.3 Method**

Photophysical analysis of the samples were carried out at the synthesis of each and every nano-components and also during each layering step. The Rh6G which was infused with the PLA core was crucial to the measurements obtained. The solutions to be analysed where placed in cuvettes and either placed in the fluorometer for emission analysis or the spectrometer as per the requirements. The absorbance range used between 300 to 800 nm because Rh6G usually has an excitation peak around 520-530 nm and this range matched with that of Rh6G. Suitable software were used to operate the instruments and to interpret the data obtained. The findings were plotted as graphs and were analysed.

### 4.9 SEM Analysis

### 4.9.1 Chemicals

- Final solution containing microcapsules

### 4.9.2 Instruments

- Hitachi Tabletop Microscope TM-1000

### 4.9.3 Method

For the sample to be analysed using SEM, the sample needs to be dry and therefore the solutions obtained in our experiment could not be used directly as such. To observe the samples in the SEM, the samples were agitated to make sure that the particles were well suspended, after which a drop of the sample was placed on a small glass slide and dried thoroughly naturally. The sample after the layer by layer assembly process was then prepared for SEM by following the above steps. Though the drying method of the sample was crude, there seemed to be no other alternate way to observe the sample in its liquid condition microscopically. When the sample was observed under the microscope however, there seemed to be no drastic effect of the drying process on the particles as clear images of the synthesised particles were seen. However, the effect the drying process may have had on the characteristics of the particles is unknown.

## 5. Results and Discussion

### 5.1 Photophysical Analyses

Absorption and emission spectra are central to photophysical analyses. To obtain a clear understanding of the analytical methods, the fundamentals must be well understood beforehand. Rh6G used in this experiment is a fluorophore. Fluorophores are photochemicals which reemit light when they are excited. Every fluorophore exhibits a unique and a characteristic adsorption/emission spectrum.

Absorbance measurements provide one with the wavelengths of the light which are absorbed by the substance. The absorption by the substance is because of the photon causing the electrons in the substance to get to an excited state. This causes a state of energy gain and this lasts only for a very few nanoseconds. It can also be said that the absorption spectra would correspond to the fluorescence emission spectra to a certain degree when approximated.

Fluorescence emission reading is dependent on the intensity as the absorbance reading. Once the substance after being hit by light returns to ground state, photon emission takes place. This causes a state of energy loss and this is usually in the form of vibrations. The light emitted here would be of lower energy when compared with that of the absorbed beam and also would have a higher wavelength.

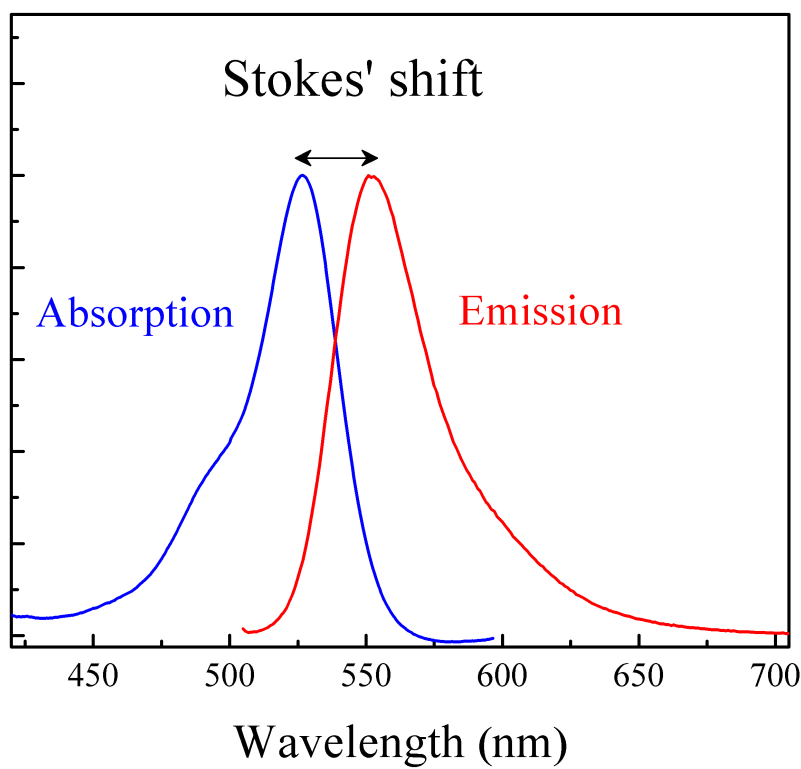

**Figure 9**: Normalised absorption spectra and emission spectra of Rh6G (Stokes' shift of around 25nm cab be observed).

Stokes' shift, which is the difference between the peak maxima of the absorbance and emission wavelengths plays a major role in the experiment. In the experiment, from the Stokes' shift one can infer if the excitation is due to the light emitted or due to the light used to excite the substance. Stokes' shift is also the reason why the emission spectra have proportionally higher wavelengths to that of the absorbance spectra as seen from the graph (Figure 9) above.

## 5.1.1 Photophysical Analysis during PLA-Rh6G synthesis

During every step of the synthesis of particles, photophysical analysis was carried out to confirm that the process was going on as intended. The mixing of Rh6G with the PLA could be observed visually by the colour of the pellets as seen in figure 10. PLA on its own is colour less or white, but the pellets below exhibit the distinct shade of orange-red corresponding to the Rh6G dye and therefore it could be confirmed that the Rh6G has indeed adhered to the PLA.

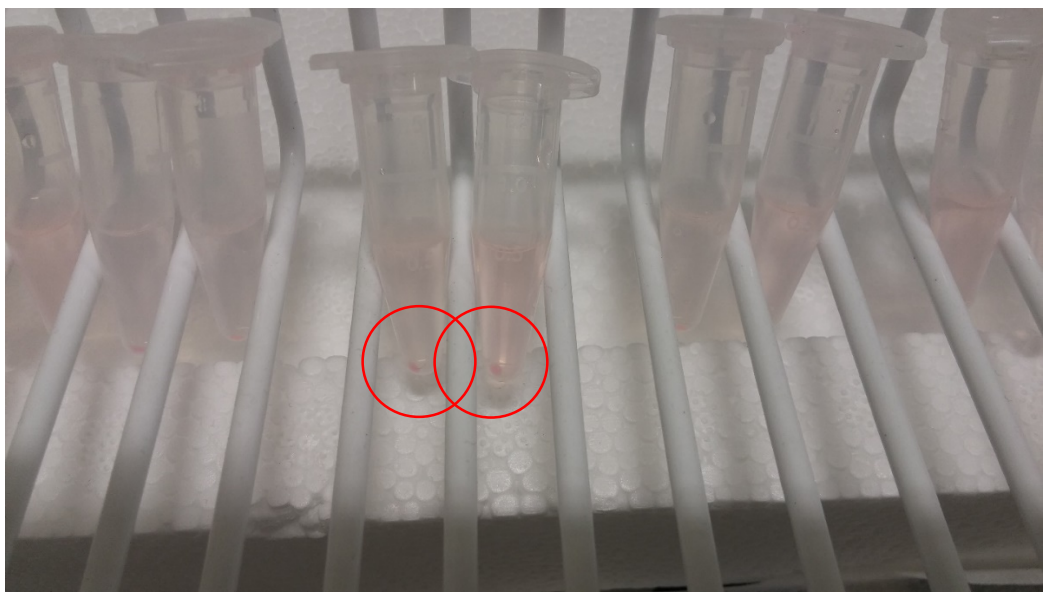

**Figure 10**: Centrifuge tubes showing the PLA-Rh6G pellets.

After the synthesis of the PLA-Rh6G particles, the supernatant was analysed. The following graph (Figure 11) was obtained through photophysical analysis.

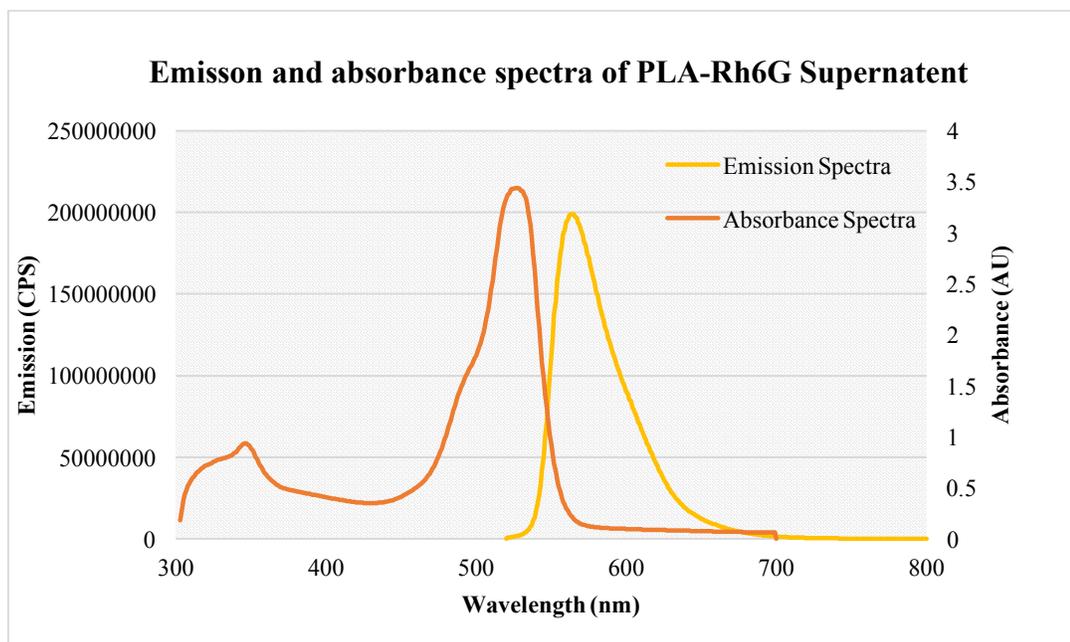

**Figure 11**: Emission and absorbance spectra of PLA-Rh6G Supernatant.

The absorbance values of the graph clearly correspond to that of the absorbance values of the Rh6G dye. A wavelength range suiting Rh6G which is between 300 and 700 nm was used. The peak of the absorbance as seen from the graph was around 525 nm. The absorbance values in the graph above correspond directly and proportionally with the intensity of colour of the supernatant solution, i.e. if the concentration of the Rh6G in the solution was increased, the colour intensity would be higher thereby producing higher absorbance values. For the supernatant in this case, at the peak of 530 nm, an absorbance value of 3.40 was obtained. Going below 340 nm wavelengths, the fluctuation of the signal occurs as the spectral region falls outside the range of the Rh6G dye as seen in the graph (Figure 11) above.

The emission spectra for Rh6G was measured from around 550nm to 800nm which is the ideal range. The peak of emission was found to be around a wavelength of 566nm. The emission intensity was deduced to be around $2 \times 10^8$. To find the Stokes' shift the difference between the wavelength peak maxima which on our case is 566-530 which gives us a value of 36 nm.

After the analysis of the supernatant, the sample containing the suspended Rh6G was then analysed as carried out previously. The following graph (Figure 12) was obtained when the data were plotted.

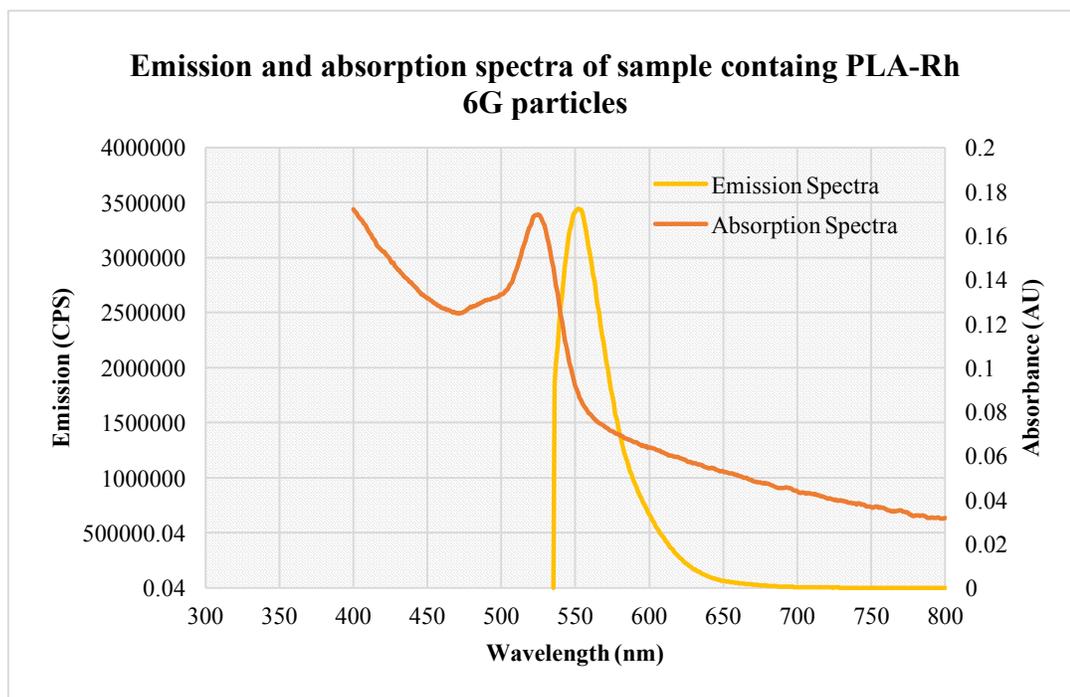

**Figure 12**: Emission and absorption spectra of sample containing PLA-Rh6G particles.

In this case, the absorbance maxima were at 528 nm and the emission peak was at 554 nm. Therefore, the Stokes' shift was 554-528 = 28 nm. The Stokes' shift here is lower than the 36 nm observed previously for the supernatant. Since the Stokes' shift is smaller a bigger overlap between the emission and absorbance spectra can be visualised. The absorbance at the peak was 0.169 nm approximately, while emission max was around $3.43 \times 10^6$.

## 5.1.2 Photophysical Analysis during Gold Nanoparticle Synthesis

The success of the procedure for gold nanospheres was marked by the characteristic colour of the solution obtained as seen below. The ruby red colour of the solution is due to the fact that the solution absorbs blue and green colours which is due to the plasmonic optical properties of the gold nanospheres. This causes the solution to emit red light which could be observed clearly.

From the spectral graph below, it could be further proven that the colour change corresponds to the nanoparticles of relative size of approximately 10nm as the graph below (Figure 13) suggests the absorption of green and blue photons at the peak which is around 530 nm.

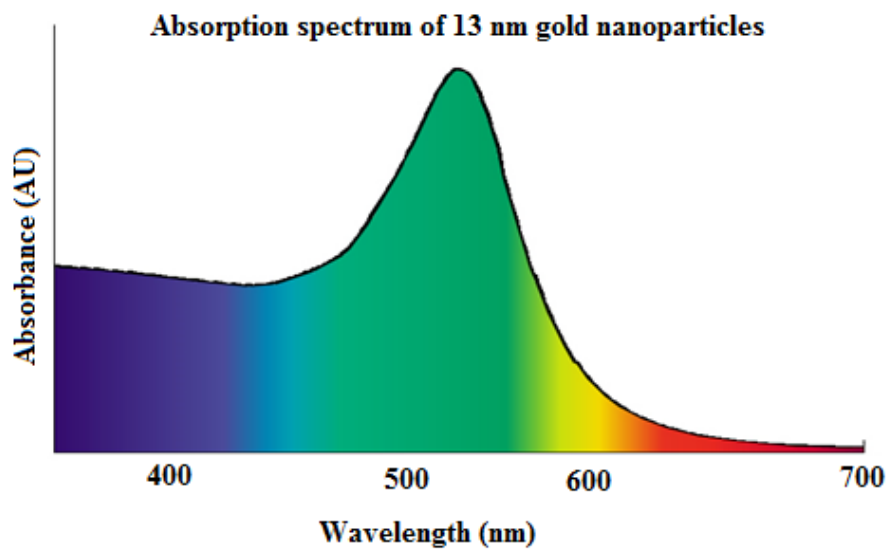

**Figure 13**: Absorption spectrum of gold nanoparticles (13 nm) [99].

Once the gold nanospheres were synthesised, the solution was analysed and the following spectral graph (Figure 14) was obtained for absorbance.

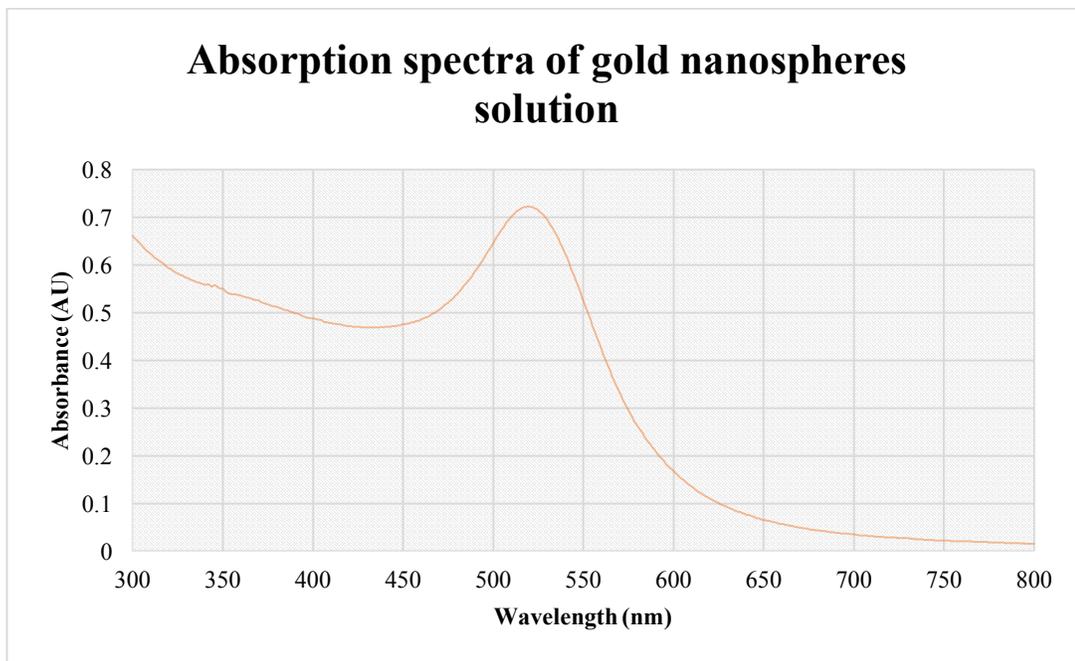

**Figure 14**: Absorption spectra of gold nanospheres solution.

The above graph displays the characteristic shape for gold nanospheres, which further supports that gold nanospheres were indeed successfully synthesised. The absorbance peak maxima occur at a wavelength of 524 nm marked by an absorbance of 0.717. This peak corresponds with the nanoparticles sized between 10 nm to 13 nm which further is in line with the intended requirement of approximately 10 nm. It can also be noted that this absorption graph corresponds with the spectral graph for the gold nanoparticles seen earlier.

### 5.1.3 Photophysical Analysis during Layer by Layer Assembly Process

### 5.1.3.1 Absorbance Measurements

Throughout the entire layer by layer assembly process, the supernatant at each step was kept aside and photophysically analysed. Absorbance measurements were taken to ensure that the Rh6G was at a minimum during each stage. Less leakage of Rh6G would mean less drug loss in the real drug delivery scenario.

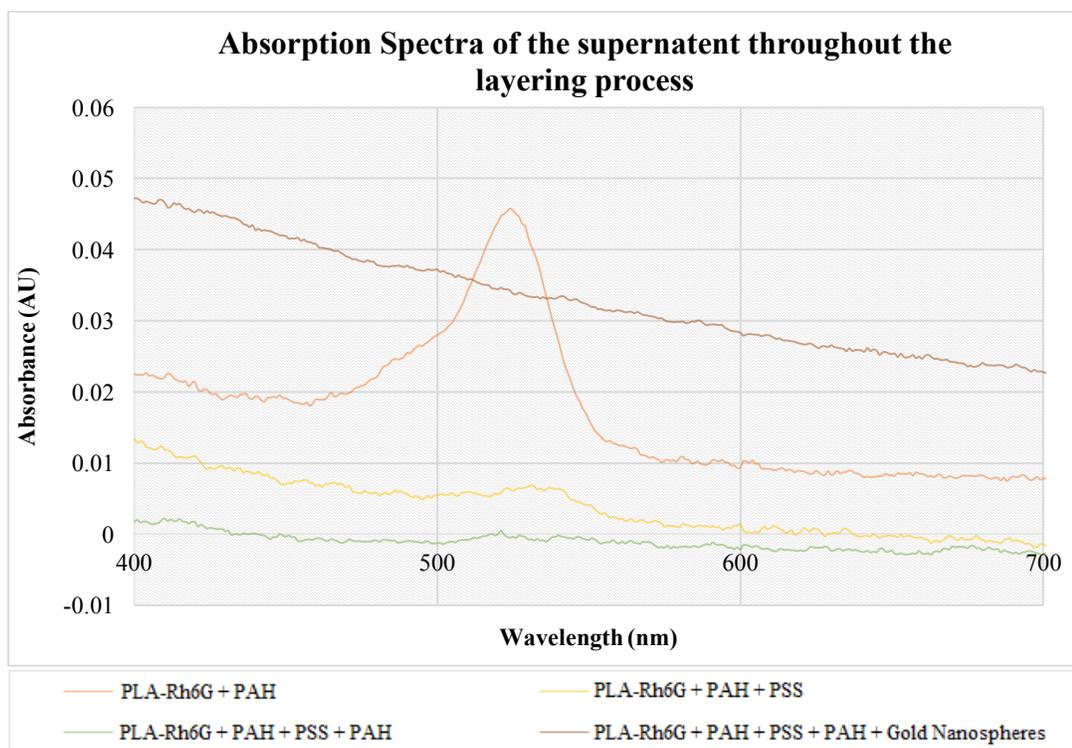

**Figure 15**: Absorption spectra of the supernatant throughout the layering process

The absorbance range was again set in favour of Rh6G which was within 400 to 700 nm. When the first PAH layering had been achieved, the supernatant was found to have an absorbance peak at the wavelength of 525 nm with an absorbance value of 0.0455. The shape of the curve seems to correspond with that of the PLA-Rh6G supernatant and this is understandable because of the possibility of the presence of Rh6G in the supernatant, since it is the first layer. Over the second and third layers, it can be seen that the detection of Rh6G seems to have significantly declined.

When the sample containing the product with the gold nanoparticles was analysed, the level of absorbance had increased and a significant shift in absorbance could be seen as in the graph above. This suggests the presence of the gold nanospheres in the top layer, and hence the huge rise in absorbance. The overall graphical data (Figure 15) support the fact of the presence of new particles in each layering step thereby indirectly supporting the synthesis of new materials.

### 5.1.3.2 Emission Measurements

The fluorescence emission spectra of the supernatant were similarly obtained and the data were plotted (Figure 16). The values were adjusted proportionally to be able to fit into a single graph thus enabling a comparative understanding.

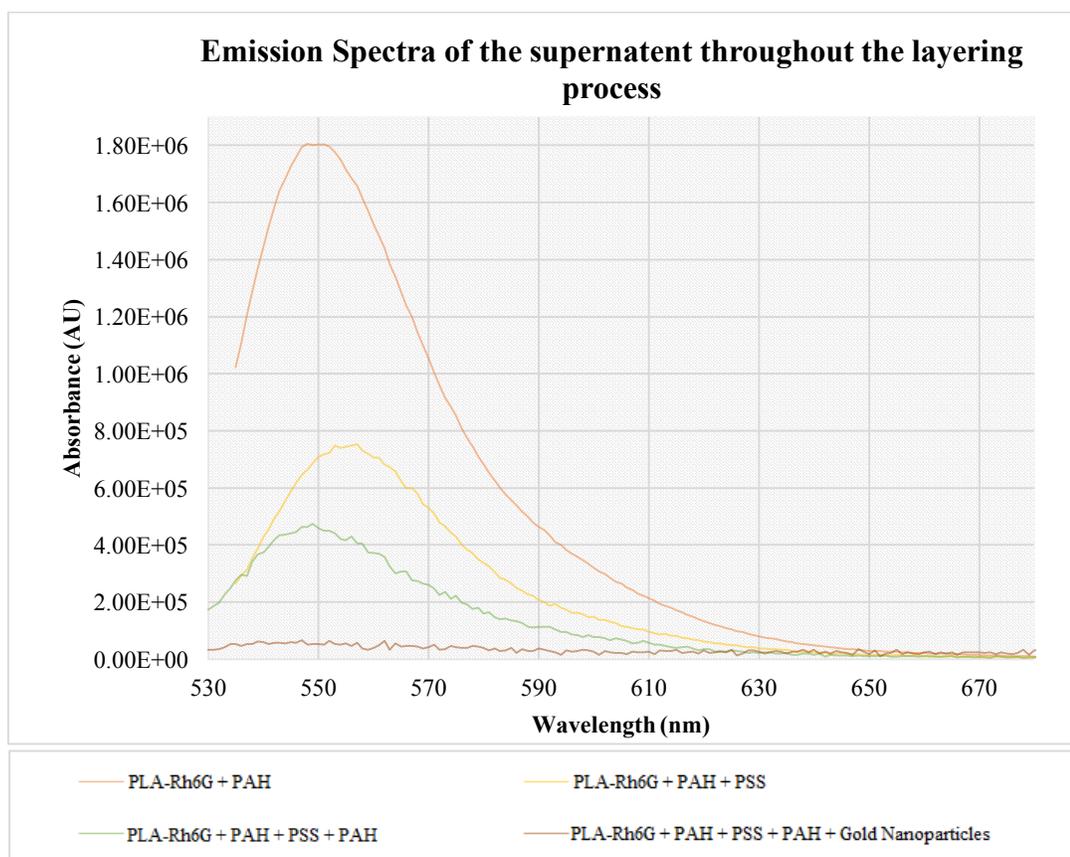

**Figure 16**: Emission Spectra of the supernatant throughout the layering process

The emission peaks for the first three layers were significantly visible in comparison to the absorbance values, where a detectible peak was only observed for the first layer. For the supernatant of the PLA-Rh6G-PAH particles the peak maximum occurs at around 551 nm. For the first PSS layer, peak max was recorded at 558 nm followed by a peak max of 550 nm for the second PAH layer. Once the gold nanosphere layer was formed, the emission curve almost flat lines with a peak max at a wavelength of 561 nm. Since the range of the peak maxima of all these supernatant remained very close, it can be inferred that the supernatant at the final steps did not have any considerable particles. This shows the efficiency of the layering process. Also it can be clearly deduced that the concentration level of Rh6G started to significantly decrease in every layering step, thus showing that the leakage of Rh6G has almost come to a halt and proving that each layer adds additional stability to the core. This finding is also in correlation with similar published studies [100]. The emission graph above together with the absorbance graph could be used to further support the efficacy of the methods that are used for the synthesis. Clear and distinct shifts between each process could be observed on the graphs which could therefore substantiate the claims that the syntheses and layering processes were efficient in the creation and the assembly of the layers.

## 5.2 SEM Analysis of the Final Microcapsules

After the final layering of the microcapsule, the solution was visualised under an SEM. Clear individual microcapsules were observed. The sizes of the obtained particles had an overall mean size of 0.3795 μm which is 379.5 nm with a standard deviation of 0.2475 μm which is 247.5 nm. Though the sizes vary slightly, the distribution seems to be consistent which shows that they are equally distributed throughout the solution and well suspended. The SEM images (Figure 17) coupled with the Spectroscopic data can be used to prove the success of synthesising the intended microcapsules.

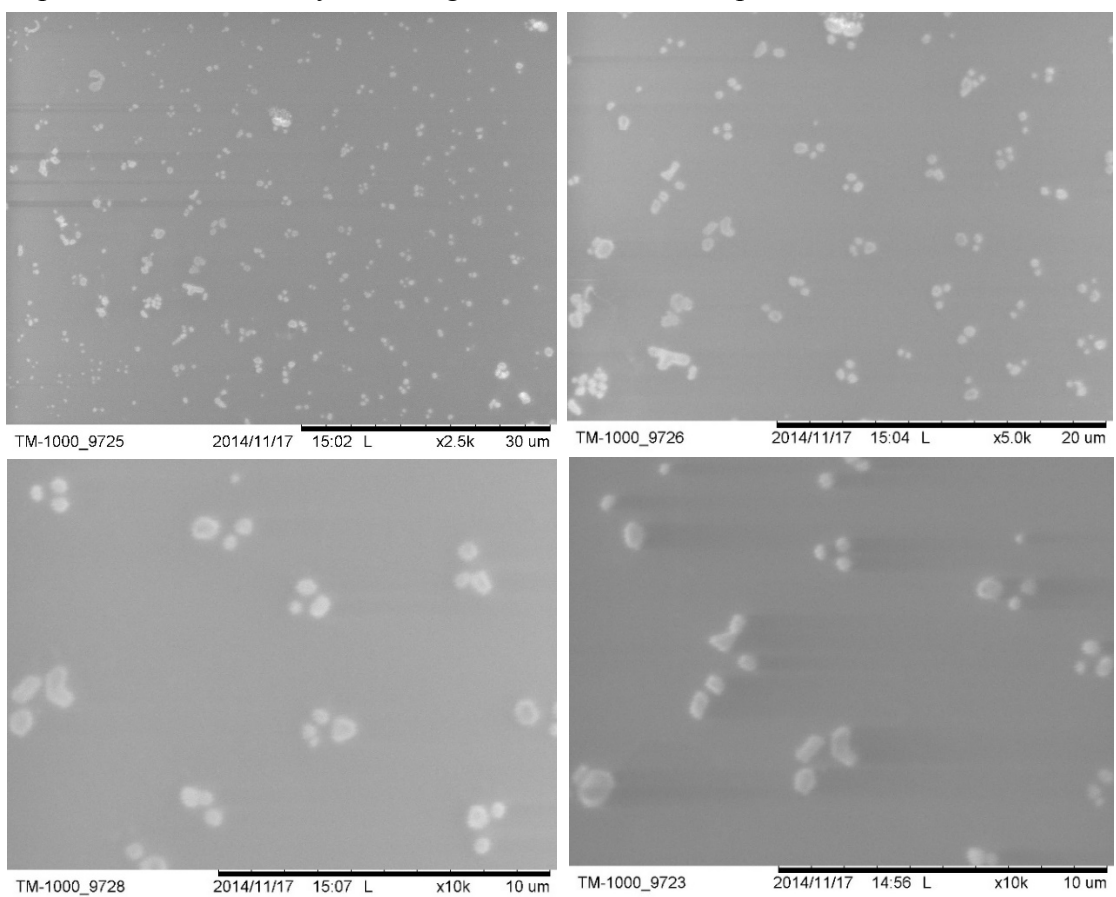

**Figure 17**: SEM images of the final PE microcapsules with the gold nanosphere layer. The two images at the top show magnifications at 2500x and 5000x each, while the bottom two shows a magnification of 10000x each.

The images at x10k magnification were analysed using ImageJ (Figure 18)

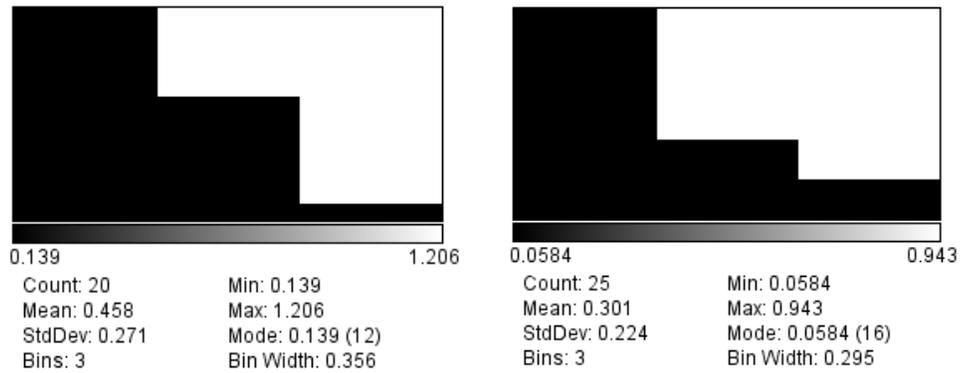

**Figure 18**: ImageJ analysis of nanoparticles (All units are in μm).

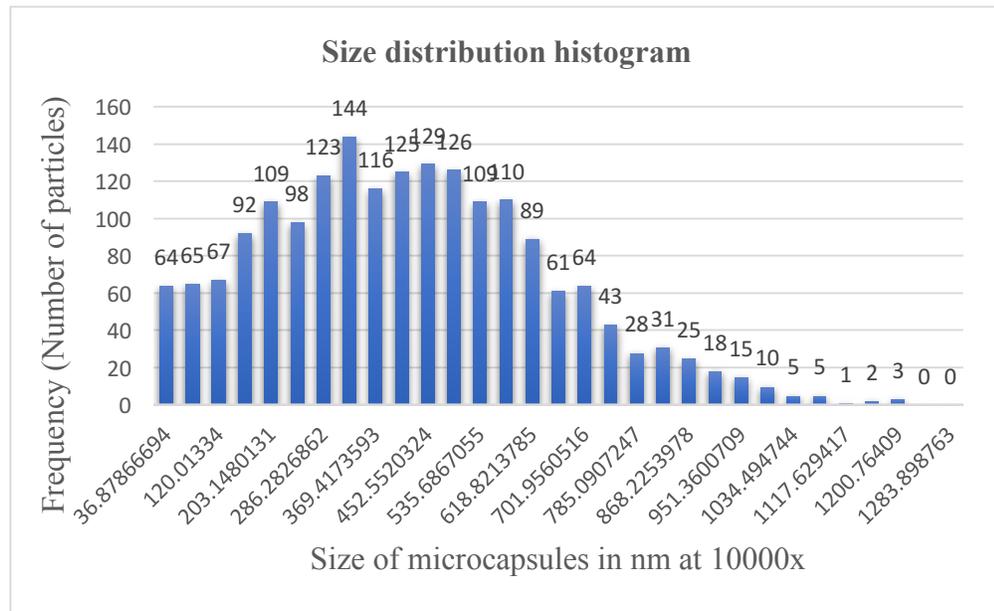

**Figure 19**: Particle size distribution histogram (All units are in nm).

The SEM data was then analysed further using excel and the above size distribution histogram (Figure 19) was obtained. The distribution of the sizes of the particles could be visualised graphically from this image. The number of particles corresponding to each sizes could be observed from the histogram above. From the histogram it could also be inferred that the particles are normally distributed. Future research should concentrate on the shapes and sizes of the particles. This would enable synthesis of particles which are of uniform dimensions and are distributed evenly.

A cause for concern was the observational process itself. Since the SEM could not be used to observe liquid samples, the sample obtained in our case had to be dried out. Therefore, it is necessary to know how the drying process could have affected the properties if the synthesised materials. The intended particles were expected to be spherical entities; the slightly irregular shapes could be blamed partially on the method of drying of the sample.

The clumping of the nanoparticles as observed in the SEM images could be also caused due to the crude drying procedure. Also since the nanoparticles have not been attached to any surface, the clumping is inevitable. However, when these nanoparticles are coated on to stents one does not need to be concerned about the clumping of the nanoparticles since they would be attached to the stents and would not be mobile thus the chances of these particles clumping together would be significantly reduced.

In future projects, the improvements on the imaging techniques could be further worked upon. Different imaging techniques such as transmission electron and atomic force microscopy could be explored to obtain better images. Also the microcapsules could be coated onto metal like gold or titanium sheets before imaging, which would enable the optimal distribution of the particles and prevent them from clumping and also from being distributed unevenly as observed here.

## 5.3 Final concept

The ultimate goal of the project was to create microcapsules which could later be coated onto stents and then eluted using novel mechanisms. The microcapsules had been obtained successfully, however further research needs to be done with regards to coating them onto stents.

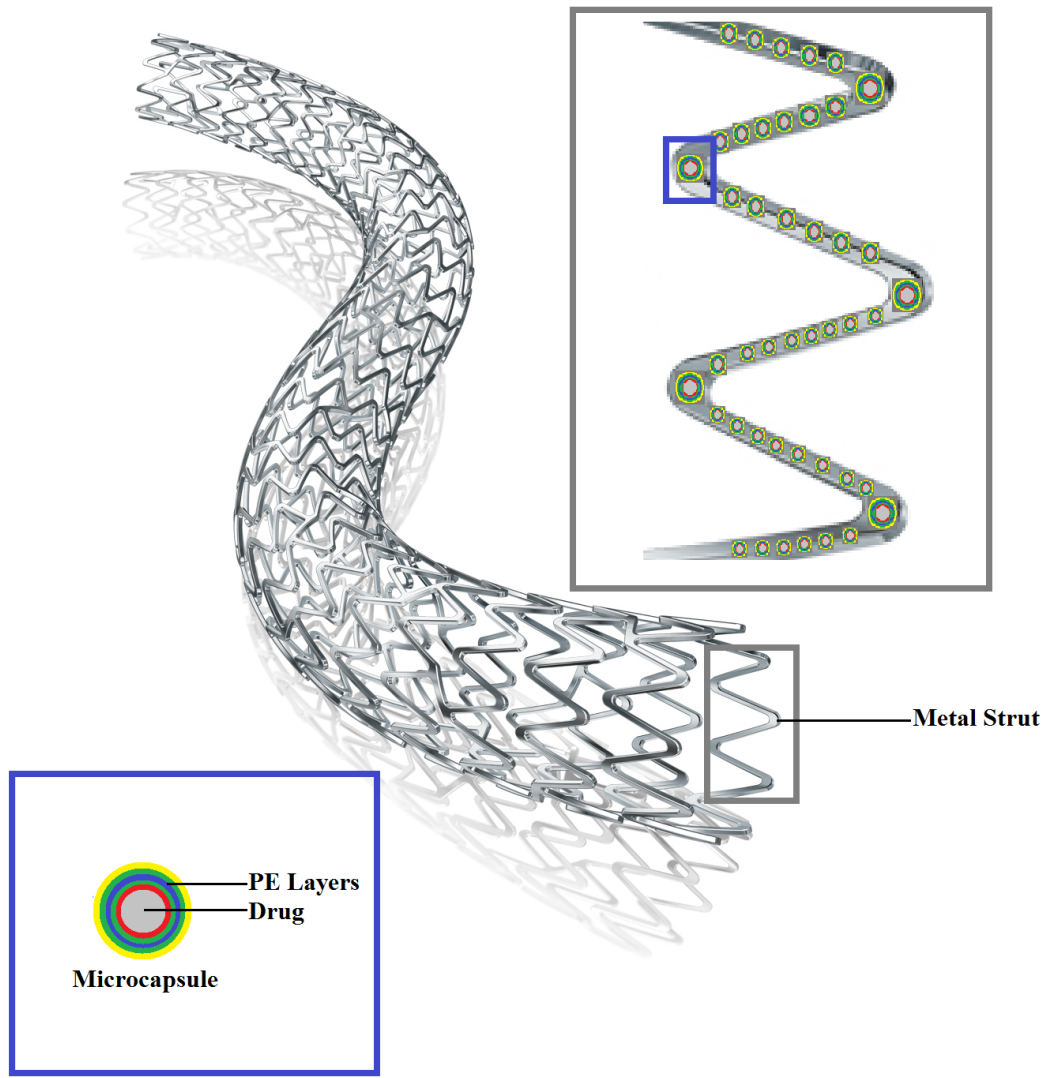

**Figure 20**: Proposed use of the obtained microcapsules in stents (Illustration only, not to scale)

As the picture above (Figure 20) explains, the microcapsules that were obtained could be coated onto stents. Theoretically, the coating could be done based on charge difference, which would mean that the stent strut would have to be charged oppositely to that of the final layer of the microcapsules. The microcapsules would then be needed to be loaded with drugs before coating it onto the stents. This can be done wither through removing the core by degrading it or by creating the capsule using the drug as the core itself. This would depend on the properties of the drug itself. If the drug is not stable enough mechanically, it would not be possible to coat layers over the said drug and in that case, using an alternate core would be followed. Once coated onto the strut of the stents, various release mechanisms could be explored. But the release mechanism of interest in our project was the use of ultrasound to elute the drugs. This could be explored in future projects to test whether the microcapsules which were created are suitable for an ultrasound triggered release system or not.

The obtained microcapsules can be used for a wide range of various purposes, the chief one being carriers for drugs. The stability and other properties of these kinds of microcapsule have been previously studied, and also the characteristics of these capsules with relation to temperature differences and materials used were studied extensively [101].

# 6. Future Research Possibilities

The experimental goal of the project was to synthesise novel microcapsules which could be used in drug delivery systems, but primarily on DESs. Studies have been carried out with regards to loading ISR treatment drugs like paclitaxel onto the shell of microcapsules. These studies found a method to control the drug concentrations in the capsule by varying the number of layers of the capsules [102]. The same study discusses a method of creating a hollow microcapsule through using a biodegradable core and later degrading it, thus providing a hollow microcapsule which could be loaded with the desired drugs.

The other main future objective was to use a new mechanism to elute the drugs from the capsules once they are coated on to the stents. The mechanism to be used would be ideally something that could control the drug release. One of the potential mechanisms that was found when exploring through research was the use of ultrasound. But research in this area should deal with the conditions which would be required to make such a system possible. High frequency ultrasound with lower power has been used widely in medicine [12]. Therefore, there has been interest in this area of using this kind of an ultrasound based stimulus to break the drug carrier to enable drug release.

But future research should focus on the various parameters which need to be tweaked to make sure that the external stimuli which might be used to trigger the drug release does not lead to any side effects. Therefore, the safety aspect of the system needs to be extensively researched before trials and potential implementation. It is also necessary to have trials with real drugs in place of Rh6G to find out the efficacy of the system in a real world scenario. Though there has been a lot of research work on treating CVDs till date, there are only few reports which deal with novel drug delivery systems such as this. Of the few that are available, some studies have even opened up some negative possibilities on the use of certain stimuli like ultrasound, which may lead to certain undesired effects on the heart like irregular rhythm [103]. Some of the studies have not taken many clinically relevant conditions into account while performing the study and this further complicates the data that have been obtained. Therefore, this area of medicine is still in its early stages and needs a lot of research for development and attaining clinically usable systems.

# 7.Conclusion

This thesis began with an aim to create an entity which would aid in the development of current drug delivery systems. The entity being the microcapsules and the drug delivery system being the DESs. The project was successful in synthesising the microcapsules through assembly of the various nano-components which were synthesised separately. It is very much hoped that the efficacy of the synthesised microcapsules for coating on stents could be studied in future research and also the potential to elute the drugs by means of ultrasound or other novel stimuli which would help one control the drug release could be explored. If future research succeeds in coating the said nanoparticles onto stents and eluting it by means of ultrasound for example, systemic drug delivery for ISR could become obsolete in the future. And although implants in the human body can be traced back to ancient times, coronary stents have been used only since the 1990s, and still they have not been yet made in such a way that they do not trigger any response form the human body. With increasing number of younger patients with CHDs who are receiving stent implants, the need for the treatment of stent related conditions such as ISR and late stent thrombosis remains highly essential, therefore stents which are coated with these microcapsules and which could remain with the stents as the patient ages could be of immense use, especially the possibility to elute the drugs at a later age of the patient avoiding the need to do an invasive procedure. Therefore, this project needs to proceed in that direction of the creation of a newer and far more advanced drug delivery system, which could lead to the development of better targeted drug delivery treatments which employ controlled drug deliveries for CHDs, and also advanced and better stents.